\documentclass{article}
\usepackage{spconf,amsmath,graphicx}
\usepackage{amssymb} 
\usepackage{bm} 
\usepackage{xcolor}
\usepackage{url}
\usepackage[]{todonotes}
\usepackage[utf8]{inputenc} 

\newcommand{\defined}{\stackrel{\text{def}}{=}}

\newcommand{\graph}{\mathcal{G}}
\newcommand{\ver}{\mathcal{V}}
\newcommand{\edg}{\mathcal{E}}
\newcommand{\adj}{\mathbf{A}}
\newcommand{\degr}{\mathbf{D}}
\newcommand{\lap}{\mathcal{L}}

\title{Improved functional MRI activation mapping in white matter through diffusion-adapted spatial filtering}

\name{David Abramian$^{a,b}$ \quad Martin Larsson$^{c}$ \quad Anders Eklund$^{a,b,d}$ \quad Hamid Behjat $^{e}$ 
\thanks{This research has been supported by the Swedish Research Council under grants 2018-06689 and 2017- 04889, the Center for Industrial Information Technology (CENIIT) at Linköping University, and the Wallenberg AI, Autonomous Systems and Software Program (WASP) funded by the Knut and Alice Wallenberg Foundation. Correspondence should be addressed to H. Behjat, e-mail: hamid.behjat@bme.lth.se.}
} 

\address{
\small $^{a}$ Department of Biomedical Engineering, Linköping University, Linköping, Sweden \\ 
\small $^{b}$ Center for Medical Image Science and Visualization, Linköping University, Linköping, Sweden\\
\small $^{c}$ Centre for Mathematical Sciences, Lund University, Lund, Sweden \\
\small $^{d}$ Department of Computer and Information Science, Linköping University, Linköping, Sweden\\
\small $^{e}$ Department of Biomedical Engineering, Lund University, Lund, Sweden
}

\begin{document}
%

\maketitle

\begin{abstract}

Brain activation mapping using functional MRI (fMRI) based on blood oxygenation level-dependent (BOLD) contrast has been conventionally focused on probing gray matter, the BOLD contrast in white matter having been generally disregarded. Recent results have provided evidence of the functional significance of the white matter BOLD signal, showing at the same time that its correlation structure is highly anisotropic, and related to the diffusion tensor in shape and orientation. This evidence suggests that conventional isotropic Gaussian filters are inadequate for denoising white matter fMRI data, since they are incapable of adapting to the complex anisotropic domain of white matter axonal connections. In this paper we explore a graph-based description of the white matter developed from diffusion MRI data, which is capable of encoding the anisotropy of the domain. Based on this representation we design localized spatial filters that adapt to white matter structure by leveraging graph signal processing principles. The performance of the proposed filtering technique is evaluated on semi-synthetic data, where it shows potential for greater sensitivity and specificity in white matter activation mapping, compared to isotropic filtering.

\end{abstract}

\begin{keywords}
functional MRI, diffusion MRI, white matter, adaptive filtering
\end{keywords}

\vspace{-0.2cm}
\section{Introduction}

Functional magnetic resonance imaging (fMRI) is a noninvasive technique that allows investigation of the activity of the brain while performing a task or at rest. Most fMRI studies report activations in the gray matter, while reports of white matter activations are relatively sparse~\cite{mazerolle2010confirming, fabri2011topographical, gawryluk2014functional, courtemanche2018detecting, huang2018voxel}. The significance of such reports is controversial \cite{gawryluk2014does}, as the sources of the blood-oxygenation level dependent (BOLD) signal in white matter are not fully understood. Despite this, significant evidence directly linking the BOLD signal in white matter to neural activity has recently been presented \cite{ding2018detection}.

The relative scarcity of reports of white matter activations can be partially explained by the anatomical and physiological differences between white and gray matter \cite{gawryluk2014does}, which suggest the potential need for different experimental designs and analysis methods to optimize detection power in each tissue type. It has been shown, for example, that increased T2-weighting improves the sensitivity to callosal activations in an interhemispheric transfer task \cite{gawryluk2009optimizing}. Therefore, we expect the development of methods geared specifically towards white matter to be required in order to investigate the functional significance of the white matter BOLD signal.

An important distinguishing feature of the BOLD signal in white matter is that it exhibits a spatial correlation structure grossly consistent with the directions of water diffusion, as measured by diffusion tensor imaging (DTI) \cite{ding2013spatio}. This correlation structure is present during rest and becomes even more pronounced under functional loading \cite{wu2017functional, ding2018detection}. As a consequence, isotropic Gaussian filtering, commonly used to increase the signal-to-noise ratio of the fMRI signal, may prove inadequate for use on the highly anisotropic white matter domain.

The anisotropy of the white matter domain makes it suited for a graph-based description. Graph signal processing methods have recently experienced significant development \cite{hammond2011wavelets,shuman2016vertex}, resulting in their successful application to the analysis of the BOLD signal in gray matter \cite{behjat2015anatomically,behjat2016signal}. In this work we propose a graph-based adaptive filtering approach for the white matter BOLD signal which incorporates information from diffusion MRI in order to adapt the filter design to the white matter mircostructure. We test our proposed filtering approach on semi-synthetic phantoms and show its increased sensitivity and specificity for activations in white matter.

\section{Methods}

\subsection{Preliminaries}

We consider undirected, connected, weighted graphs $\graph=(\ver,\edg,\adj)$, where $\ver$ is a set of vertices with $|\ver| = N_v$, $\edg$ a set of edges connecting pairs $(i,j)$ of vertices, and $\adj$ is an adjacency matrix whose nonzero elements $a_{i,j}$ represent the weight of edges $(i,j) \in \edg$. The diagonal degree matrix, denoted $\degr$, associated to $\adj$ is defined with elements $d_{i,i} = \sum_j a_{i,j}$. 

The normalized Laplacian matrix of $\graph$ can then be obtained as $ \lap = \mathbf{I} - \degr^{-1/2} \adj \degr^{-1/2}$. As $\lap$ is a real symmetric matrix, it can be diagonalized, resulting in a set of $N_v$ real non-negative eigenvalues which define the graph spectrum $\mathbf{\Lambda}$, i.e., $ \mathbf{\Lambda} = \{0=\lambda_1 \leq \lambda_2 \ldots \leq \lambda_{N_v} \defined \lambda_{\text{max}} \leq 2 \}$. The associated eigenvectors, denoted $\{ \boldsymbol{\chi}_l \}_{l=1}^{N_v}$, form an orthonormal basis spanning the $\ell^2(\graph)$ space. Here $\ell^2(\graph)$ represents the Hilbert space of all square-integrable graph signals $\bm{f} : \ver \to \mathbb{R}$ defined on the vertex set $\ver$. A graph signal can therefore be seen as a vector $\bm{f} \in \ell^2(\graph)$ whose $n$-th component represent the signal value at the $n$-th vertex of $\graph$. A graph signal $\bm{f}$ can be filtered with spectral kernel $\mathcal{K}(\lambda)$ as 
\begin{equation}
	\label{eq:directApproach}
    \left( F_{\bm{k}} \bm{f} \right)[m]
    = \sum_{l=1}^{N_v} \hat{\bm{k}}[l] \hat{\bm{f}}[l] \bm{\chi}_l[m] ,
\end{equation}
where $\hat{\bm{k}}$ is the sampled version of $\mathcal{K}(\lambda)$ obtained as $\hat{\bm{k}}[l] = \mathcal{K}(\lambda_l), l = 1, \dots, N_v$ and $\hat{\bm{f}}$ is the graph Fourier transform of $\bm{f}$ (see \cite{ortega2018graph} for an introduction to graph signal processing).

\subsection{White matter graph construction}

We define a white matter graph $\graph^{(\text{WM})}$ as a graph whose vertex set $\ver^{(\text{WM})}$ consists of all the white matter voxels in a brain volume (approximately $250$k for the data used). The set of edges $\edg^{(\text{WM})}$ is defined such that each voxel is connected to every voxel in a specified neighborhood, which we define as a $5 \times 5 \times 5$ region, resulting in at most $124$ neighbors for each white matter voxel.

We define the weight of the edge that joins a pair of voxels $(i,j)$ in analogy to Iturria-Medina et al. \cite{iturria2007characterizing} and Sotiropoulos et al. \cite{sotiropoulos2010brain}. Let $O_{i}(\omega)$ denote the diffusion orientation distribution function (ODF) associated to voxel $v_{i}$, with its coordinate center being the center of voxel $v_{i}$. Let $\mathcal{N}_i$ denote the set of vertices in $ \ver^{(\text{WM})}$ that are adjacent to vertex $i$; i.e., $\mathcal{N}_i : \{k\in \ver^{(\text{WM})} \vert (i,k) \in \edg^{(\text{WM})}\}$. For any two vertices $i, j \in \ver^{(\text{WM})}$, let $\vec{r}_{i,j}$ denote the vector pointing from the center of vertex $i$ to the center of vertex $j$, and define
\begin{align}
    p(i, \vec{r}_{i,j}) = \int_{\Omega_{i,j}} O_{i}^{n}(\omega)  d\omega , 
    \label{eq:pDiff}
\end{align}
where $n\in \mathbb{Z}^{+}$ is a desired power factor to sharpen the ODFs and $\Omega_{i,j}$ denotes a solid angle of $4\pi/98$ around $\vec{r}_{i,j}$ subtended at the center of voxel $v_{i}$. It is desirable to sharpen the ODFs since they generally manifest only slight variations between directions of strong and weak diffusion. Using (\ref{eq:pDiff}), the weight between vertices $i$ and $j$, denoted $w_{i,j}$ is defined as
\begin{equation}
    w_{i,j} = \frac{p(i, \vec{r}_{i,j})}{C_{i}} + \frac{p(j, \vec{r}_{j,i})}{C_{j}} ,
\end{equation}
where $ C_{k} = 2 \max_{l \in \mathcal{N}_{k}} p(k, \vec{r}_{k,l})$. The normalization factor ensures having $w_{i,j} \in [0,1], \quad \forall i \in \ver^{(\text{WM})}, \forall j\in \mathcal{N}_{i}$.

Using diffusion MRI, $O_{i}(\omega)$ is estimated at a discrete set of directions $\{\vec{r}_{k}\}_{k=1}^{N_{o}}$ from the center of an ODF, denoted $\{O_{i,k}\}_{k=1}^{N_{o}}$, which can be used to approximate $p(i, \vec{r}_{i,j})$ as a sum
\begin{align}
    p(i, \vec{r}_{i,j}) \approx \frac{4\pi}{N_{o}} \sum_{k\in \mathcal{D}_{i,j}} O_{i,k}^{n}, 
\end{align}
where $\mathcal{D}_{i,j}:\{k \: \vert \: \vec{r}_{k} \in \Omega_{i,j}\}$. The resulting edge weights amount to a measure of coherence in the directions of diffusion at neighboring voxels, with high weights associated with highly coherent diffusion and vice versa. 


\subsection{Graph filter definition}
Conventional filters defined within the Euclidean domain, such as Gaussian filters, encompass a shift invariant impulse response. In contrast, vertex realizations of graph filters are \emph{shift-variant}, leading to a unique spatial realization of the filter at each graph vertex. This property, combined with the weighting scheme in the proposed graph definition, results in graph filters adapted to the microstructure of white matter.

Due to the lack of shift-invariance, it is convenient, and conventional, to specify graph filters in the graph spectral domain. In this work, we design and leverage smoothing filters associated with a heat kernel spectral profile, defined by
\begin{equation}
    \quad \mathcal{K}(\lambda) = e^{-\tau \lambda},\quad \forall \lambda \in [0,\lambda_{\text{max}}],
\end{equation}
where $\tau$ is a free parameter determining the spatial extent of the filter. In particular, $\mathcal{K}(\lambda) \in L^2(\graph)$, where $L^2(\graph)$ denotes the Hilbert space of all square-integrable ${\mathcal{K}: [0,\lambda_{\max}] \to \mathbb{R}^+}$. Although such filters are roughly analogous in shape to the Gaussian filters typically used for fMRI data analysis, there is no direct equivalence between them.

\begin{figure*}[tbp]
    \centering
    \includegraphics[width=0.98\textwidth]{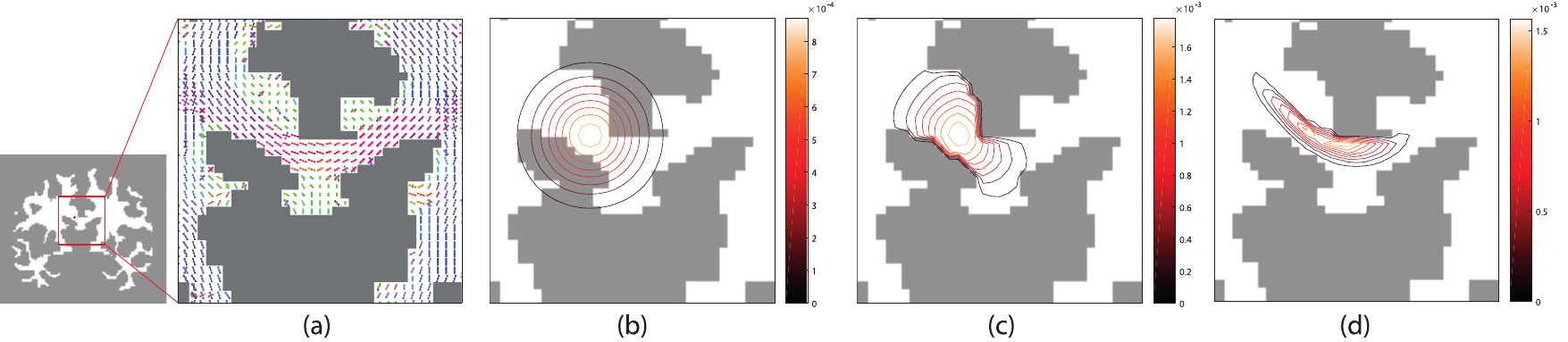}
    \caption{(a) Main orientation of ODFs manifesting white matter fiber directions at a coronal slice in the vicinity of the corpus callosum. (b)-(d) Comparison of the contour profiles of three spatial filters, with different extents of adaptation to local anatomy, localized at a point within the corpus callosum marked with a red dot in (a). (b) An isotropic Gaussian filter adapts neither to local tissue morphology nor the underlying white matter structure. (c) A filter restricted to white matter adapts to the local tissue morphology, but not to the underlying white matter structure. (d) A filter defined on the proposed graph adapts both to the tissue morphology and the underlying white matter structure.}
    \label{fig:atoms}
\end{figure*}

\subsection{Graph filtering }

Implementing graph filtering as in (\ref{eq:directApproach}) requires calculation of all the eigenvectors of the Laplacian matrix $\{\bm{\chi}_l\}_{l=1}^{N_v}$, which is practically infeasible for larger graphs such as those proposed here. Instead, we use a fast approximation algorithm \cite{hammond2011wavelets}. Let $\mathcal{P} \in L^2(\graph)$ be a polynomial approximation of kernel $\mathcal{K}(\lambda)$. For a graph signal $\bm{f}$, its filtering with kernel $\mathcal{K}$ can be found using $\mathcal{P}$ as
\begin{equation} 
	\label{eq:poly_approx}
	\tilde{\bm{c}}_{\mathcal{K}}
    = \sum_{l=1}^{N_v} \mathcal{P}(\lambda_l) \hat{\bm{f}}[l] \bm{\chi}_l 
    = \mathcal{P}(\lap) \sum_{l=1}^{N_v} \hat{\bm{f}}[l] \bm{\chi}_l 
    = \mathcal{P}(\lap) \bm{f} , 
\end{equation}
where $\tilde{\bm{c}}_{\mathcal{K}} \in \ell^2(\graph)$ with $\tilde{\bm{c}}_{\mathcal{K}}[m]=\left( F_{\bm{k}} \bm{f} \right)[m]$; in the last equality we exploit: 
$ \lap \bm{\chi}_l = \lambda_l \bm{\chi}_l \Rightarrow \mathcal{P}(\lap) \bm{\chi}_l = \mathcal{P}(\lambda_l) \bm{\chi}_l$. 

This approximation has the benefit that it does not require the explicit calculation of the eigenvectors. Instead, a polynomial of the Laplacian matrix is applied to the signal, which can be efficiently implemented with matrix-vector multiplication. Similarly to \cite{hammond2011wavelets}, we leverage a truncated Chebyshev polynomial expansion approximation of $\mathcal{K}(\lambda)$, as it has the benefit of approximating a minimax polynomial, minimizing an upper bound on the approximation error.

\vspace{-0.2cm}
\section{Data}

The MRI data used in this work was acquired from the Human Connectome Project \cite{van2013wu}. Structural images, parcellations, and diffusion data from the ``$100$ Unrelated Subjects" subset of the $1200$ subject release were used. Data from five of the subjects was excluded from the analysis due to incomplete white matter coverage of the diffusion data. Additionally, fMRI data from a motor task from one subject was used.

To evaluate the performance of the proposed graph-based filtering method we created a set of semi-synthetic phantoms, consisting of synthetic activation patterns generated on the basis of tractography results from real diffusion data. Tractography was performed using generalized Q-sampling imaging \cite{yeh2010generalized} in DSI Studio \cite{dsistudio}. Synthetic activation patterns were created from individual streamlines by diffusing an activation along the length of the streamline, starting from a random point. 

Activation patterns from $100$ streamlines were combined to produce a single ground truth activation pattern for each subject. An example of such an activation pattern can be seen in Figure~\ref{fig:phantom}(a). Time-series phantoms for all subjects were then created by convolving the ground truth activation pattern with a regressor corresponding to a block fMRI paradigm. The resulting time-series were contaminated with additive white Gaussian noise of $\sigma=1$.

\section{Results}

\vspace{-0.1cm}
\subsection{Spatial adaptivity of graph filters}
\vspace{-0.1cm}




Figure~\ref{fig:atoms} illustrates the differences in spatial adaptivity from various filter definitions. Isotropic Gaussian filters adapt neither to the local tissue morphology nor the underlying white matter microstructure. They transcend tissue boundaries, integrating signal components associated to white matter as well as the adjacent gray matter/CSF. Moreover, the filters integrate signal components from multiple unrelated white matter tracts.

By restricting the analysis to the white matter tissue, the filters do not transcend tissue boundaries and thus only combine signal components from white matter. However, the filters can not differentiate signal components from unrelated white matter tracts.

Filters defined on the proposed white matter graphs adapt both to the local tissue morphology and the underlying microstructure. Their spatial profile closely matches the anisotropic diffusion orientation manifested by the ODFs.
    
\vspace{-0.1cm}
\subsection{Semi-synthetic phantom results}
\vspace{-0.1cm}

\begin{figure}[tbp]
    \centering
    \includegraphics[width=0.95\columnwidth]{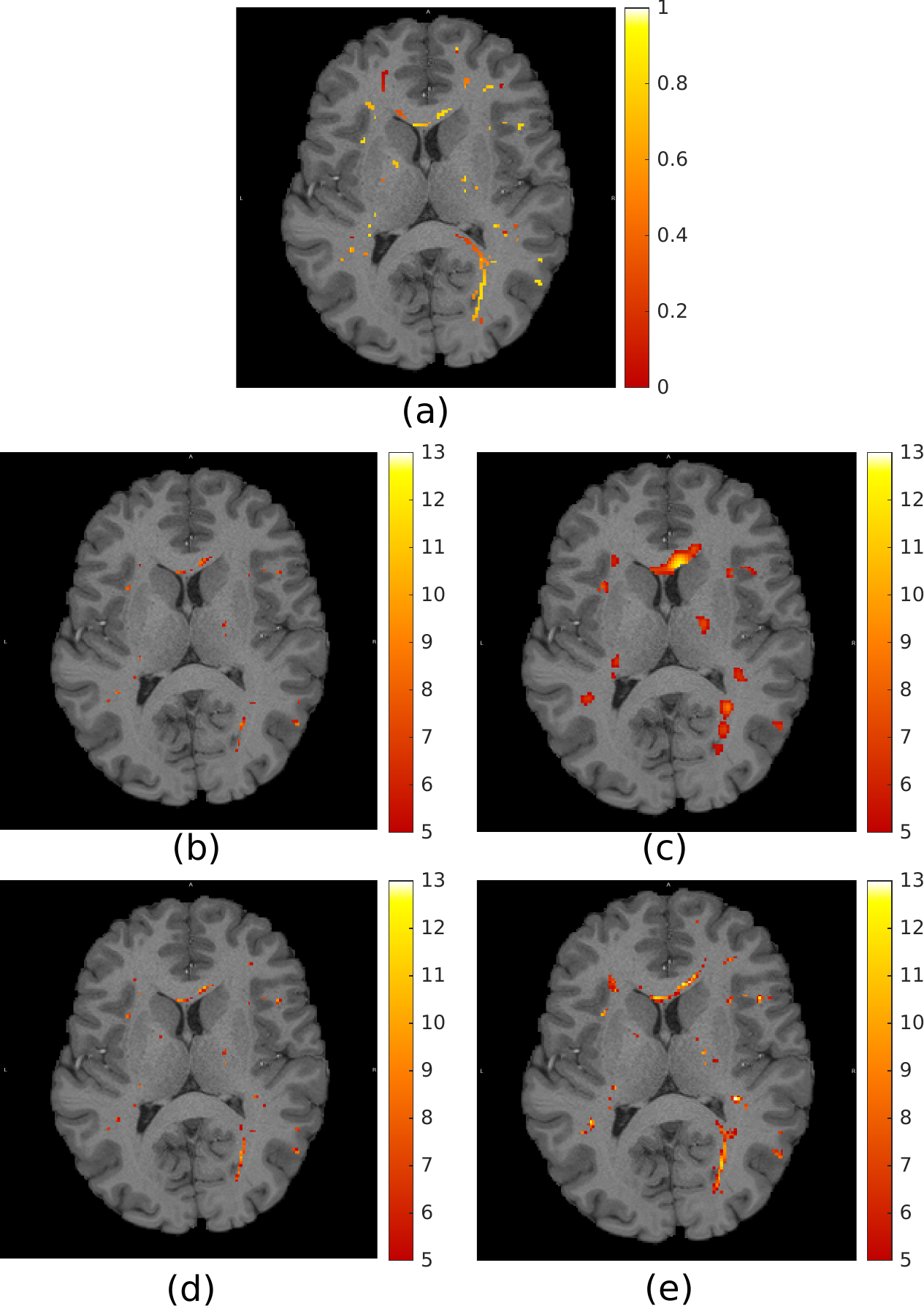}
    \caption{(a) Example synthetic activation pattern. Dot-shaped activations extend linearly in the plane orthogonal to the image. (b)-(e) t-maps obtained from analysis conducted using: (b)-(c) isotropic Gaussian filtering, $\text{FWHM}=2$mm and $6$mm respectively; (d)-(e) graph filtering, $\tau=1.3$ and $3.3$ respectively. All t-maps thresholded at $t=5$ and overlaid on the T1 image of the corresponding subject.}
    \label{fig:phantom}
\end{figure}

The $95$ semi-synthetic time-series phantoms were filtered using both isotropic Gaussian filtering and the proposed graph filtering approach. The resulting filtered volumes were subjected to a standard GLM-based activation mapping using the SPM toolbox. Figure~\ref{fig:phantom} shows a comparison of the results obtained by using the various filtering methods. For small filter sizes, isotropic Gaussian filtering is capable of detecting the subtle shapes of the activations, but with reduced sensitivity compared to graph filtering. More activations are detected with larger filter sizes, but at the cost of diminished specificity. In contrast, the shapes of the activations detected through graph filtering are consistent across filter sizes.

\begin{figure}[tbp]
    \centering
    \includegraphics[width=0.99\columnwidth]{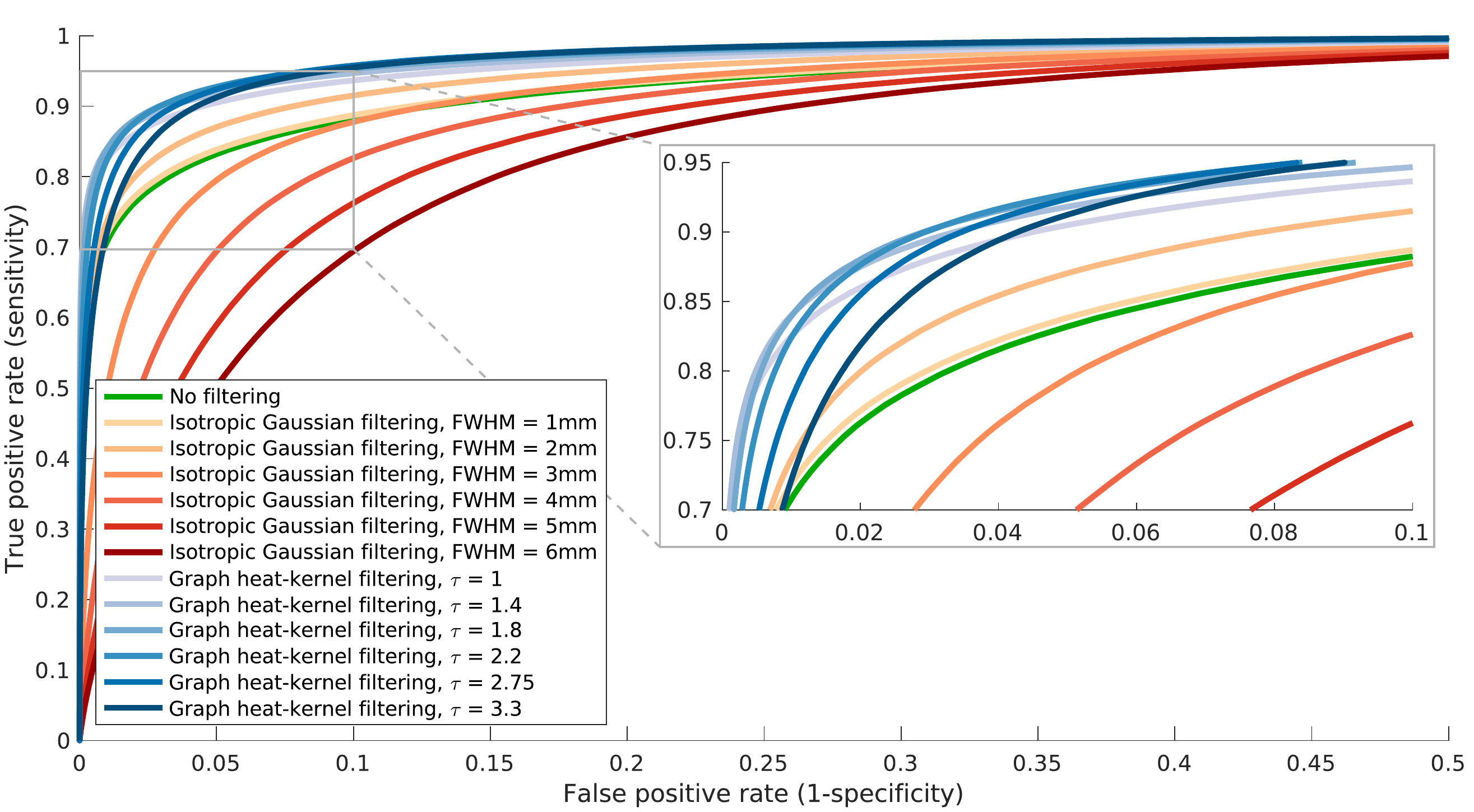}
    \caption{Average ROC curves from $95$ semi-synthetic phantoms. The proposed graph-filtering approach shows increased sensitivity and specificity across filter sizes.}
    \label{fig:ROCs}
\end{figure}

The t-maps from all phantoms were thresholded at multiple levels and compared with the available ground truth activation patterns in order to produce ROC curves, which were then averaged across subjects. Figure~\ref{fig:ROCs} shows the average ROC curves for a variety of filter sizes. The best performance was achieved with Gaussian filters of $\text{FWHM}=2$mm and graph filters of $\tau=1.4$. The proposed filtering approach results in increased sensitivity and specificity across all tested filter sizes. Importantly, the performance of Gaussian filters of $\text{FWHM}>2$mm is worse than without the application of any filtering.

\vspace{-0.1cm}
\subsection{Task fMRI results}
\vspace{-0.1cm}

\begin{figure}[b]
    \centering
    \includegraphics[width=0.9\columnwidth]{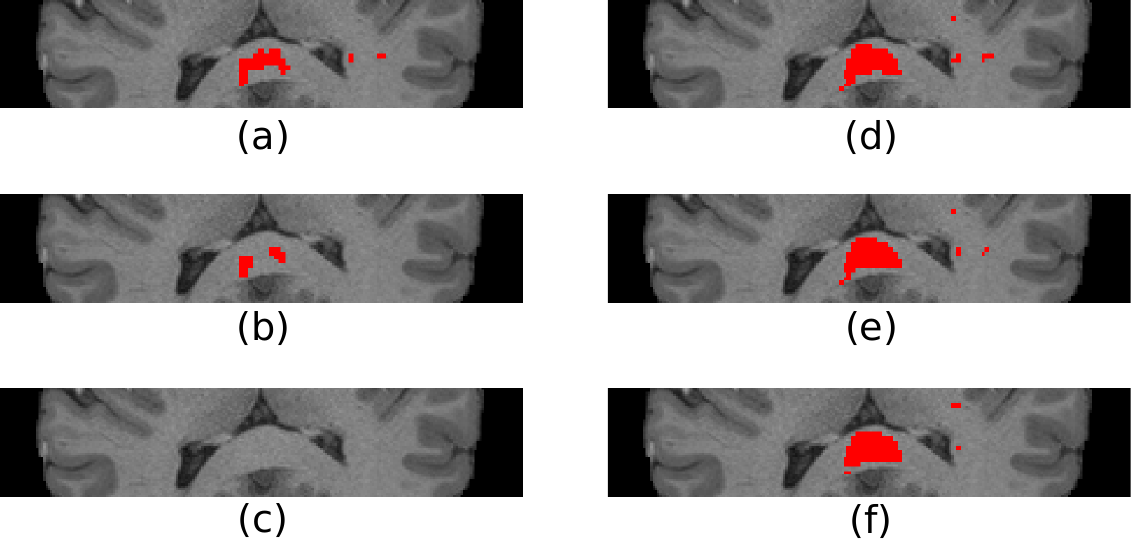}
    \caption{Single subject activations in the corpus callosum from a left hand motor task using: (a)-(c) isotropic Gaussian filtering, $\text{FWHM}=2$mm, $4$mm and $6$mm respectively. (d)-(e) graph filtering, $\tau = 1.4$, $2.2$ and $3.3$ respectively. All t-maps thresholded at 5\% FDR.}
    \label{fig:realdata}
\end{figure}

As a proof of concept, results from a single HCP subject performing a motor task are presented; see Figure~\ref{fig:realdata}. A substantially greater extent of activations was detected in the corpus callosum from fMRI data filtered using the proposed approach than from data filtered with isotropic Gaussian smoothing. Similarly to the results obtained on phantoms, the detection performance for isotropic Gaussian filtering deteriorates with increasing filter sizes as a result of the increased ratio of extraneous signal mixed with the signal of interest. In contrast, the proposed filters show more homogeneous performance across filter sizes, as they consistently adapt to the shape of the underlying signal.

\vspace{-0.05cm}
\section{Conclusion}
\vspace{-0.25cm}

We have proposed a novel graph-based approach for filtering white matter fMRI data. The design enables constructing shift-variant spatial filters that adapt to the underlying white matter structure, which enable revealing fine-grained, anisotropic activity patterns. Results on semi-synthetic data showed the potential of the proposed approach to enable improved specificity and sensitivity in white matter activity detection, compared to the use of isotropic Gaussian filtering. Our future work will focus on testing the proposed approach on a larger cohort of subjects, to probe white matter activations across different functional loading tasks as well as to investigate resting state BOLD signal fluctuations in white matter.

\clearpage
\newpage
\bibliographystyle{IEEEbib}
\bibliography{refs}

\end{document}